\documentstyle[12pt]{article}
\begin{document}
\pagestyle{plain}
\setcounter{page}{1}
\baselineskip16pt

\begin{titlepage}

\begin{flushright}
PUPT-1555\\
hep-ph/9508279
\end{flushright}
\vspace{20 mm}

\begin{center}
{\huge A Simple Description of Strange Dibaryons}

\vspace{5mm}

{\huge in the Skyrme Model}
\end{center}

\vspace{10 mm}

\begin{center}
{\large Igor R.\ Klebanov and Karl M.\ Westerberg }

\vspace{3mm}

Joseph Henry Laboratories\\
Princeton University\\
Princeton, New Jersey 08544

\end{center}

\vspace{2cm}

\begin{center}
{\large Abstract}
\end{center}

We study strange dibaryons based on the SU(2)-embedded $B=2$ toroidal
soliton. Treating the excursions of the soliton into strange directions
as small rigid oscillations, we obtain a good approximation to the
bound state approach. We calculate the dibaryon
mass formula to order $1/N$ and find that the doubly strange $I=J=0$
dibaryon is bound by about $90\,{\rm MeV}$.

\vspace{2cm}
\begin{flushleft}
August 1995
\end{flushleft}
\end{titlepage}
\newpage
\renewcommand{\baselinestretch}{1.1}  


\newcommand{\AAA}[3]{A_{#1} A_{#2} A_{#3}}
\newcommand{\commAA}[2]{\left[A_{#1},A_{#2}\right]}
\newcommand{\DK}[1]{D_{#1}K}
\newcommand{\dK}[1]{\partial_{#1}K}
\newcommand{\hc}[1]{{#1}^\dagger}
\newcommand{\leftparens}[1]{\left( #1 \right.}
\newcommand{\mAA}[2]{A_{#1} A_{#2}}
\newcommand{\mtxIIIxIII}[4]{\parens{
   \begin{array}{c|c}
      {\displaystyle \vphantom{\underline{#1}} #1}
         & {\displaystyle \vphantom{\underline{#2}} #2} \\ \hline
      {\displaystyle \vphantom{\sqrt{#3}} #3}
         & {\displaystyle \vphantom{\sqrt{#4}} #4}
   \end{array}
}}
\newcommand{\noparens}[1]{#1}
\newcommand{\parens}[1]{\left( #1 \right)}
\newcommand{\power}[2]{{#1}^{#2}}
\newcommand{\rightparens}[1]{\left. #1 \right)}
\newcommand{\taudot}[1]{#1 \cdot \vec{\tau}}
\newcommand{\trAAA}[3]{tr\parens{\AAA{#1}{#2}{#3}}}
\newcommand{\trAA}[2]{tr\parens{\mAA{#1}{#2}}}

\renewcommand{\epsilon}{\varepsilon}
\newcommand{\SU}[1]{\mbox{${\rm SU}(#1)$}}
\newcommand{\WZ}{{\rm WZ}}
\newcommand{\QCD}{{\rm QCD}}
\newcommand{\disk}{{\rm D}}
\newcommand{\grad}{\nabla}
\newcommand{\MeV}{{\rm MeV}}
\newcommand{\rightvbar}[1]{\left. #1 \right|}
\newcommand{\tr}{\mathop{\rm tr}}
\newcommand{\sym}{\mathop{\rm sym}}
\newcommand{\id}{1}
\newcommand{\mtxIIxI}[2]{\parens{
   \begin{array}{c}
      #1 \\
      #2
   \end{array}
}}
\newcommand{\mtxIIxII}[4]{\parens{
   \begin{array}{cc}
      #1 & #2 \\
      #3 & #4
   \end{array}
}}

\newcommand{\sA}{{\cal A}}
\newcommand{\sB}{{\cal B}}
\newcommand{\sD}{{\cal D}}
\newcommand{\sL}{{\cal L}}
\newcommand{\sM}{{\cal M}}

\newcommand{\va}{\vec{a}}
\newcommand{\vb}{\vec{b}}
\newcommand{\vn}{\vec{n}}
\newcommand{\vv}{\vec{v}}
\newcommand{\vw}{\vec{w}}
\newcommand{\vx}{\vec{x}}
\newcommand{\vJ}{\vec{J}}
\newcommand{\valpha}{\vec{\alpha}}
\newcommand{\dalpha}{\dot{\valpha}}
\newcommand{\vtau}{\vec{\tau}}

\newcommand{\bc}{\bar{c}}
\newcommand{\bs}{\bar{s}}
\newcommand{\bomega}{\bar{\omega}}

\newcommand{\unitr}{\widehat{r}}
\newcommand{\unittheta}{\widehat{\theta}}
\newcommand{\unitphi}{\widehat{\phi}}

\newcommand{\dr}{{\rm\bf d}r}
\newcommand{\dtheta}{{\rm\bf d}\theta}
\newcommand{\dphi}{{\rm\bf d}\phi}
\newcommand{\dt}{{\rm\bf d}t}

The existence of stable multibaryon states with vanishing hypercharge,
although not yet experimentally confirmed, remains an exciting
possibility. The simplest of such states, the doubly strange dibaryon
$H$, was first conjectured to exist in \cite{jaffe}.
On the basis of an MIT
bag model calculation, its mass was predicted to be $m_H=2150\,\MeV$,
well below the $\Lambda\Lambda$ threshold. This means that all strong
decays of $H$ are forbidden, and it is expected to have a long lifetime
typical of weak decays. Experimental detection of
doubly strange dibaryons is a subtle
matter, and there are some remarkable ongoing efforts
in this direction \cite{alan}.
In the meantime it is important to sharpen the
theoretical understanding of dibaryons
by resorting to other available models of low-energy QCD.

One viable alternative to the quark models
is the Skyrme model \cite{sky,bal,wit}
where baryons are identified with solitons of the non-linear meson
action, which is usually taken to be
\begin{equation}
S = N\,S_{\WZ}+\int d^4 x\left [
\frac{f_\pi^2}{16} \tr(\partial_\mu \hc{U} \partial^\mu U)
   + \frac{1}{32e^2} \tr[\partial_\mu U \hc{U},\partial_\nu U \hc{U}]^2
   + \frac{f_\pi^2}{16} \tr \sM(U + \hc{U} - 2)\right ] \label{eqn1}
\end{equation}
with $U(\vx,t)\in \SU{3}$. $N$ is the number of colors,
and the Wess-Zumino term was first determined in \cite{wit}.
$$\sM=\parens{\begin{array}{ccc}
 m_\pi^2 & & \\
 & m_\pi^2 & \\
 & & 2m_K^2-m_\pi^2
\end{array}}$$
is proportional to the quark
mass matrix. While $m_\pi=138\,\MeV$ is small and is often neglected,
the effects of $m_K=495\,\MeV$ are significant.
With the standard fit to the nucleon and
$\Delta$ masses obtained with this Lagrangian \cite{adn},
the parameters are assumed to be
$f_\pi=108\,\MeV$ and $e=4.84$.

After a remarkable success in describing the properties of non-strange
baryons, based on the SU(2) collective coordinate quantization of the
Skyrme hedgehog \cite{adn,anw}, there has been a number
of attempts to model
strange dibaryons. The first interesting idea appeared in \cite{bala}
where an SO(3) imbedded soliton of baryon number $B=2$ was
constructed. In the limit of vanishing $m_\pi$ and $m_K$,
the classical mass of the SO(3) soliton was found to be
$1658\,\MeV= 1.92 M_h$.
The classical mass of the Skyrme hedgehog
is $M_h= 863\,\MeV$.
Thus, in the chiral limit, the SO(3) soliton is stable against
decay into two $B=1$ states.
Its collective coordinate quantization leads to SU(3) multiplets of
zero triality and $H$, the SU(3) singlet,
is the lightest state \cite{jk}.
An important feature of the SO(3) soliton, however,
is that it extends significantly into the strange directions of SU(3).
As $m_K$ is dialed to its physical value, which is appreciable,
the SO(3) symmetry crucial to the soliton's existence is destroyed.
A perturbative estimate shows that inclusion of $m_K$ pushes
$M_{{\rm SO}(3)}$ well above $2 M_h$, destroying its
classical stability \cite{KY}.
Thus, the
dibaryon states constructed by the collective coordinate quantization
of the SO(3) soliton, which were found to be stable in the chiral limit,
may not survive the breaking of SU(3). This question requires further
study.

In this paper we focus on
another $B=2$ soliton solution, which is embedded
entirely in the light SU(2) subgroup of SU(3), and is based
on the cylindrically symmetric ansatz \cite{cyl},
\begin{equation}
U_{B=2}(\vx)
   = \mtxIIIxIII{e^{iF(r, \theta)\,\hat n\cdot \vtau}}{0}{\hc{0}}{1} \ ,
\qquad \hat n=\parens{\begin{array}{c}
   \sin\Theta(r,\theta) \cos2\phi \\
   \sin\Theta(r,\theta) \sin2\phi \\
   \cos\Theta(r,\theta)
\end{array}}\,.
\label{toroid}
\end{equation}
The classical mass of
this soliton has no dependence on $m_K$. By numerically relaxing
the two unknown functions with the boundary conditions
$$F(r=0)=\pi, \ F(r=\infty) = 0, \ \Theta(\theta=0)=0,
   \ \Theta(\theta=\pi/2)=\pi/2,$$
the classical mass was found \cite{cyl}
to be $M_{B=2} \approx 1660\,\MeV$, which is considerably smaller than
the mass of the SO(3) soliton evaluated with $m_K=495\,\MeV$.
Constructions of dibaryon states based on the soliton (\ref{toroid})
may be found in the literature \cite{dib,kunz,tsw}.
In \cite{dib}, an SU(3) collective coordinate quantization
of $U_{B=2}$ was carried out, and certain stable dibaryon states were
predicted. While these predictions are interesting, it should be
noted that a similar approach to the octet and the
decuplet of baryons has not
been particularly successful because of problems with the breaking of
SU(3) \cite{aneesh}.
There exists another approach to strangeness \cite{CK}, however, which
makes no explicit mention of SU(3) multiplets but nevertheless
appears to be more successful quantitatively \cite{CK,bs1988b}.
In this approach, hyperons are modeled by bound states of
kaons and SU(2) solitons \cite{CK}.

The bound state approach was first applied
to strange dibaryons in \cite{kunz}.
To simplify calculations, the ansatz (\ref{toroid}) was restricted to
$ F(r,\theta)= F(r), \ \Theta(r, \theta) =\theta $.
This restriction effectively forces the soliton into a spherically
symmetric shape and pushes its classical energy above $2 M_h$.
While an interesting qualitative picture emerged, no definitive
assessment of dibaryon stability could be made.
More recently, an improved study of bound state dibaryons was made
in \cite{tsw}. In this paper, a better (although not minimum
energy) $B=2$ soliton ansatz was used, and kaon modes were studied in its
background.  The lightest dibaryon was predicted
to be bound by about $35\,\MeV$. This calculation, however, did not
take full account of kaon-kaon interactions, which are difficult
to include in the bound state approach, but are expected
to be particularly important in this system.

In this letter, we propose a  new description
of the Skyrme model dibaryons which we call
the rigid oscillator approach
and which is intermediate between the collective coordinate and the
bound state approaches. As in the collective coordinate quantization,
we allow only the rigid motions of the soliton.
As in the bound state approach,
we expand in the deviations of the soliton into strange directions,
which we denote by $D$.
Since $D$ turns out to be of order
$1/\sqrt N$, this
generates a systematic $1/N$ expansion. The suppression of
the strange deviations is related to the fact that, in the quark
model language, we are dealing with states consisting of $2N-2$ light
quarks and only 2 strange quarks.
While increasing $m_K$ further
reduces the strange deviations, our methods work for any
$m_K$.\footnote{This is a Skyrme model manifestation of the general
fact that for large $N$ baryons, flavor SU(3) breaking is large even for
small $m_K$ \cite{Klebanov,djm}.}
An advantage of the rigid oscillator approach is its simplicity:
it provides analytic formulae for various quantities in terms of
$m_K$, thus serving as a good physical guide to the bound state
calculations. Furthermore, it is not hard to include all terms
quartic in $D$, which are necessary for complete order $1/N$
calculations. The rigid oscillator approach has been applied to
the $B=1$ sector in \cite{Klebanov,Kaplan&Klebanov,wk},
giving a reasonable approximation
to the bound state approach. In \cite{west}, it was further noted that
the rigid oscillator approximation gets better with increasing baryon
density and beyond some critical density becomes identical to
the bound state approach. We may expect, therefore, that
the rigid oscillator approach will work better for the denser
$B=2$ soliton than for the Skyrme hedgehog.

Let us start by reviewing the rigid oscillator approach to quantization
of the $B=1$ Skyrme hedgehog,
$$U_h(\vx)
   = \mtxIIIxIII{e^{iF(r)\,\unitr \cdot \vtau}}{0}{\hc{0}}{1} \ .$$
While in \cite{Klebanov,Kaplan&Klebanov,wk}
the approximation with $m_\pi=0$ was considered, here
we extend it to include the pion mass. We consider only the rigid
motions of the soliton,
$$U(\vx,t) = \sA(t) U_h(\vx) \hc{\sA(t)}\ .$$
To separate the SU(2) rotations from the deviations into
strange directions, we write \cite{Kaplan&Klebanov}
\begin{equation}
\sA(t) =
\mtxIIIxIII{A(t)}{0}{\hc{0}}{1}S(t)\label{anso} \ ,
\end{equation}
where $A(t) \in \SU{2}$,
and
\begin{equation}
S(t) = \exp i \sum_{a=4}^{7} d^a \lambda_a
  = \exp i \sD \ , \label{anst}
\end{equation}
where
\begin{equation}
\sD = \mtxIIIxIII{0}{\sqrt 2 D}{\sqrt 2\hc{D}}{0}\ ,\qquad\qquad
   D = \frac{1}{\sqrt2} \mtxIIxI{d^4 - id^5}{d^6 - id^7} \ .
\label{ansth}\end{equation}
In calculating to order $N^0$, we may neglect the dynamics
of $A(t)$ and treat the strange deviations in the harmonic
approximation.
This leads to the following effective Lagrangian,
\begin{equation}
L =
-M_h
+4 \Phi_1 \hc{\dot{D}} \dot{D}
+i {N\over 2} \parens{\hc{D} \dot{D}-\hc{\dot{D}} D}
-\Gamma_1 (m_K^2-m_\pi^2) \hc{D} D \ .
  \label{leff}
\end{equation}
The quantities $\Phi_1$ and $\Gamma_1$, whose integral
expressions are given in \cite{wk}, may be evaluated numerically,
$$ \Phi_1\approx 0.00186/\MeV\ ,\qquad\qquad
\Gamma_1\approx 0.00398/\MeV \ .
$$
Canonical quantization of (\ref{leff}) leads to the following Hamiltonian,
\begin{equation}
H=
M_h
+\frac{1}{4 \Phi_1} \hc{\Pi} \Pi
-i {\frac{N}{8 \Phi_1} \parens{\hc{D} \Pi-\hc{\Pi} D}}
+\parens{\Gamma_1 (m_K^2-m_\pi^2)+\frac{\power{N}{2}}{16 \Phi_1}} \hc{D} D
\ .
\end{equation}
The order $N$ piece of the Hamiltonian is the classical
ground state energy $M_h$.  The order $1$ piece includes the terms
quadratic in $D$ and $\Pi$, and thus may be diagonalized exactly
using creation and annihilation operators
$$D^i= \frac{1}{\sqrt{N\mu_1}} (a^i+ (\hc{b})^i)\ , \qquad
  \Pi^i= \frac{\sqrt{N\mu_1}}{2i} (a^i- (\hc{b})^i)\ ,$$
where
$$\mu_1 = \sqrt{1+16(m_K^2-m_\pi^2)\Gamma_1 \Phi_1/N^2}\ .
$$
The operators $\hc{a}$ ($\hc{b}$) may be thought of as
creation operators for constituent strange quarks (anti-quarks).
The normal-ordered Hamiltonian to order $1$ is given by
$$H = M_h + \omega_1 \hc{a}a + \bomega_1 \hc{b}b\ ; \qquad
 \omega_1 = \frac{N}{8\Phi_1}(\mu_1 - 1)\ , \qquad
 \bomega_1 = \frac{N}{8\Phi_1}(\mu_1 + 1)\ .$$
Thus, replacing a light quark with a strange quark (anti-quark)
costs energy $\omega_1$ ($\bomega_1$).  Note that for
$m_K=m_\pi$, $\omega_1$ vanishes thereby restoring the
original \SU{3} symmetry (it costs no energy to replace a $u$ or
$d$ quark with an $s$ quark) but that $\bomega_1$ tends to a
rather large value, $N / (4\Phi_1)$.
The Wess-Zumino term, which acts as magnetic field in the
$D$--$\hc{D}$ plane, breaks the
$s \leftrightarrow \bs$ symmetry.
Using the calculated values of $\Phi_1$ and $\Gamma_1$, we
find that
$\omega_1\approx 200\,\MeV$. This is a reasonable estimate of
the difference between the strange and the light quark constituent masses.
It is also a good upper bound on the similar quantity (the lowest
mode energy) found in the bound state approach to strangeness.
We conclude that the rigid oscillator approach is a sound approximation
and proceed to apply it to the $B=2$ soliton.

Since the $B=2$ soliton is less symmetric (its flavor rotations are
in general different from spatial rotations),
we consider rigid rotations both in the flavor space and in
the real space,
$$U(\vx,t) = \sA(t) U_{B=2}(R(t)\vx) \hc{\sA(t)}\ ,$$
together with the ansatz of eqs.\ (\ref{anso}--\ref{ansth}).
To order 1 we obtain an effective Lagrangian very similar to that
in the $B=1$ sector,
\begin{equation}
L =
-M_{B=2}
+4 \Phi_2 \hc{\dot{D}} \dot{D}
+i N \parens{\hc{D} \dot{D}-\hc{\dot{D}} D}
-\Gamma_2 (m_K^2-m_\pi^2) \hc{D} D \ .
  \label{newleff}
\end{equation}
The only modifications are the extra factor of 2 in the Wess-Zumino term,
and the new integral expressions \cite{dib}
$$\Gamma_2 ={f_\pi^2\over 2} \int d^3 x\, \bigl (1-\cos F(r,\theta)\bigr)
\ ,$$
$$\Phi_2={f_\pi^2 \over 8}\int d^3 x\, (1-\cos F)\left [
1+{1\over e^2 f_\pi^2}
   \left\{(FF) +(\Theta\Theta)\sin^2 F+ 4\sin^2 F {\sin^2 \Theta\over
   r^2 \sin^2\theta}\right \} \right ]\ ,
$$
where $(FF)=(\partial F/\partial r)^2+ (\partial F/\partial\theta)^2/r^2$.
Numerical evaluation of the integrals in the $B=2$ soliton background
yields
$$ \Phi_2\approx 0.0038/\MeV\ ,\qquad\qquad
\Gamma_2\approx 0.0079/\MeV \ .
$$
Canonical quantization of (\ref{newleff}) leads to the Hamiltonian
\begin{equation}
H=
M_{B=2}
+\frac{1}{4 \Phi_2} \hc{\Pi} \Pi
-i {\frac{N}{4 \Phi_2} \parens{\hc{D} \Pi-\hc{\Pi} D}}
+\parens{\Gamma_2 (m_K^2-m_\pi^2)+\frac{\power{N}{2}}{4 \Phi_2}} \hc{D} D
\ .
\end{equation}
Diagonalizing it as before, we find
$$H = M_{B=2} + \omega_2 \hc{a}a + \bomega_2 \hc{b}b\ ; \qquad
 \omega_2 = \frac{N}{4\Phi_2}(\mu_2 - 1)\ , \qquad
 \bomega_2 = \frac{N}{4\Phi_2}(\mu_2 + 1)\ ,$$
where
$$\mu_2 = \sqrt{1+4(m_K^2-m_\pi^2)\Gamma_2\Phi_2/N^2}\ .
$$
Substituting the numbers we obtain $\omega_2\approx 198\,\MeV$.
This is only $2\,\MeV$ less
than $\omega_1$, the corresponding quantity in the $B=1$ calculation.
In the bound state calculations \cite{kunz,tsw}, the kaon mode energies
in the $B=2$ and $B=1$ backgrounds were also close to each other.

In order to calculate the $1/N$ corrections we need to include
the terms in the Lagrangian which are quartic in $D$ or involve
the soliton angular velocities.
For $B=1$ this was done in \cite{wk}, and the correction to the Hamiltonian
was found to be
\begin{equation}
{1\over 2\Omega_1} \left ( (\vec I^{bf})^2 + 2 c_1 \vec I^{bf}\cdot \vec T
+\bar c_1 \vec T^2\right ) \ ,
\end{equation}
\begin{equation}
c = 1 - \frac{\Omega}{2\mu\Phi}(\mu - 1)\ ,\qquad \qquad
  \bc = 1 - \frac{\Omega}{\mu^2 \Phi}(\mu - 1)\ ,\label{cbc}
\end{equation}
where we omit the subscript 1 throughout the last equation.
$\vec I^{bf}$, the isospin relative to the body fixed axes, is the momentum
conjugate to $\vec \alpha$ which is defined by
$$\hc{A}\dot{A} = \frac12 i \dalpha\cdot\vtau \ .$$
For the non-exotic states containing no $b$-quanta,
$\vec T= {1\over 2} a^\dagger \vec\tau a$.
The $\Lambda$-particle has $\vec I= \vec I^{bf}=0$ and $T=1/2$, so that
the order $1/N$ correction to its mass is
$$ \delta m_\Lambda = {3 \bar c_1\over 8 \Omega_1} \approx 23\,\MeV \ ,
$$
where we use $\Omega_1=0.00514/\MeV$. To order $1/N$, our methods give
$m_\Lambda\approx 1086\,\MeV$, which is close to its physical value of
$1115\,\MeV$.

In extending the calculation to the $B=2$ soliton, in addition to the
SU(2) angular velocity $\dot{\vec\alpha}$ we include the spatial
angular velocity $\dot{\vec\beta}$, which is also of order $1/N$.
The complete order $1/N$ correction to the Lagrangian (\ref{newleff}) is
\begin{eqnarray*}
&\displaystyle\delta L= {1\over 2} \Omega_2\sum_{j=1}^2 (\dot\alpha_j+
i D^\dagger\tau_j \dot D- i\dot D^\dagger\tau_j D)^2
+ {1\over 2} \lambda (\dot\alpha_3- 2 \dot\beta_3
+i D^\dagger\tau_3 \dot D- i\dot D^\dagger\tau_3 D)^2& \\
&\displaystyle\mbox{}+ {1\over 2} \tilde\lambda\sum_{j=1}^2 (\dot\beta_j)^2
- N \hc{D} \,\taudot{\dot{\vec{\alpha}}}\, D
-{\frac{2}{3}i N \parens{\hc{D} \dot{D}-\hc{\dot{D}} D}} \hc{D} D
-2i \Phi_2 \parens{\hc{D} \,\taudot{\dot{\vec{\alpha}}}\,
\dot{D}-\hc{\dot{D}} \,\taudot{\dot{\vec{\alpha}}}\, D}& \\
&\displaystyle\mbox{}-\frac{8}{3} \Phi_2 (\hc{D} D) (\hc{\dot{D}} \dot{D})
+\frac{2}{3} \Phi_2 \power{\parens{\hc{D} \dot{D}+\hc{\dot{D}} D}}{2}
+2 \Phi_2 \power{\parens{\hc{D} \dot{D}-\hc{\dot{D}} D}}{2}
+\frac{2}{3} \Gamma_2 (m_K^2-m_\pi^2) \power{\parens{\hc{D} D}}{2}\ ,&
\end{eqnarray*}
where the additional moments of inertia are numerically found to be
$$ \Omega_2\approx 0.0106/\MeV\ ,\qquad
\lambda \approx 0.0072/\MeV\ ,\qquad
\tilde\lambda \approx 0.016/\MeV\ .
$$
The isospin and angular momentum relative to the
body fixed axes are
$$ I^{bf}_i ={\partial L\over \partial \dot\alpha_i} \ ,\qquad
J^{bf}_i ={\partial L\over \partial \dot\beta_i} \ ,
$$
and we find a constraint
$$ J^{bf}_3 =-2 \left (I^{bf}_3 +T_3 \right ) \ .
$$
The calculation of the Hamiltonian is lengthy, but the result is
quite simple,
\begin{equation}
\delta H_{1/N}=
{1\over 2\Omega_2} \left ( (\vec I^{bf})^2 + 2 c_2 \vec I^{bf}\cdot \vec T
+\bar c_2 \vec T^2\right ) + {1\over 2\tilde\lambda}
(\vec J^{bf})^2 + \left ({1\over 8\lambda}-{1\over 8\Omega_2}-
{1\over 2\tilde\lambda}\right ) (J^{bf}_3)^2 \ ,
\end{equation}
where $c_2$ and $\bar c_2$ are given by (\ref{cbc})
with subscripts 2 throughout.
Note that, unlike in \cite{kunz,tsw}, there is no explicit $T_3^2$ term
in the Hamiltonian. This greatly simplifies the diagonalization
of the Hamiltonian.

Using the identities,
$$ \vec T^2 = \frac14 (\hc{a}a)^2+ \frac12 \hc{a}a= {S\over 2}\left(
{S\over 2}-1\right)\ ,$$
$(\vec J^{bf})^2= J(J+1)$, and $(\vec I^{bf})^2= I(I+1)$,
we have
$$\delta M_{1/N} =\frac{1}{2\Omega_2}\left\{ c_2 K(K+1) +
(1-c_2) I(I+1)+ {\bc_2 - c_2\over 4} (S^2-2 S) \right \}
$$
$$
+ {1\over 2\tilde\lambda}J(J+1)
+ \left ({1\over 8\lambda}-{1\over 8\Omega_2}-
{1\over 2\tilde\lambda}\right ) (J^{bf}_3)^2
\ ,$$
where $\vec K=\vec I^{bf}+\vec T$.
The significance of the quantum number $K$ is that
$K(K+1)$ is equal, up to an additive constant, to the quadratic
Casimir of the SU(3) representation which emerges in the
$m_K= m_\pi$ limit. In such a limit, $c_2=\bc_2=1$ and we
recover the mass formula of the SU(3) collective coordinate
quantization. Furthermore, states with $K=0$ merge into
a $(0, N)$ SU(3) multiplet, states with $K=1$ --- into
$(2, N-1)$, states with $K=2$ --- into $(4, N-2)$, etc.

As discussed in \cite{tsw}, not all possible quantum numbers are allowed.
There are certain constraints which arise due to special symmetries of
the soliton. They work, in effect, to insure the correct statistics
of the overall wave function.
Consider, for instance, dibaryons with $S=-2$ and
$J=0$ (which immediately implies $T=1$ and $J^{bf}_3=0$).
Then one finds that the allowed $(I, K)$ quantum
numbers are $(0, 1)$, $(1, 1)$, $(2, 1)$, $(2, 3)$, etc.,
while the forbidden ones are $(1, 0)$, $(1, 2)$, $(2, 2)$,
etc.
In the SU(3) limit, the allowed $J=0$ multiplets are
$(2, N-1)$, $(6, N-3)$, etc. Remarkably, these multiplets
are also allowed in the quark model \cite{jaffe}.
In general, the states constructed from the $B=2$, SU(2) soliton
together with the states constructed from the SO(3) soliton
appear to cover all the quantum numbers found in the quark model.

Among the
$S=-2$ dibaryons that we constructed, the lightest one has
quantum numbers $I=J=J^{bf}_3=0$ and $K=1$.
Its mass to order $1/N$ is
\begin{equation}
M= M_{B=2}+ 2 \omega_2+{\bar c_2\over \Omega_2}\approx 2084\,\MeV \ .
\label{mass}\end{equation}
The binding energy with respect to $\Lambda\Lambda$,
calculated entirely within our approach, is
$(66+4+18=88)\,\MeV$, where we separated the classical, order 1, and order
$1/N$ contributions. The lightest $S=-2$ dibaryon we found
is not the $H$ because it originates from the $(2, N-1)$ SU(3)
multiplet, not from the singlet. For arbitrary $N$ it has strangeness
$S=-2$, while the $H$ has $S=-2 N/3$. Thus, for large
$N$ the breaking of SU(3) makes the state that we found lighter
than the $H$, which appeared in the SO(3) dibaryon quantization.
Whether this conclusion may be extrapolated to $N=3$ is an open
question. We believe, however, that the quantum numbers predicted
by the Skyrme model are in accord with the quark model for any $N$,
and that we have presented a good description of strange dibaryons for
sufficiently large $N$.

How well can we trust the binding energy
of $88\,\MeV$? Indeed, no model of low
energy QCD is perfect. In our approach to
the dibaryon, quantization of other
soliton modes, better treatment of the SU(3) breaking, etc., could
affect the numbers appreciably. Nevertheless, our calculation is attractive
for its simplicity, and it gives a good physical picture of both
$B=1$ and $B=2$ states. Our results strongly suggest that a tightly bound
doubly strange dibaryon is theoretically natural.

\section*{Acknowledgements}

We are grateful to Alan Schwartz for interesting discussions.
This work was supported in part by DOE grant DE-FG02-91ER40671,
the NSF Presidential Young Investigator Award PHY-9157482,
James S. McDonnell Foundation grant No. 91-48,
and an A. P. Sloan Foundation Research Fellowship.

\end{document}